\documentclass[12pt, preprint]{aastex}

\def\spit{{\it Spitzer}}

\def\70um{70~\micron}
\def\160um{160~\micron}
\def\24um{24~\micron}
\def\um{\micron}
\def\ld{L_{\rm dust}/L_{\star}}

\def\MEarth{M_\oplus}

\def\gapp{\lower 3pt\hbox{${\buildrel > \over \sim}$}\ }
\def\lapp{\lower 3pt\hbox{${\buildrel < \over \sim}$}\ }
\def\proptosim{\lower 3pt\hbox{${\buildrel \propto \over \sim}$}\ }

\def\arcsec{$^{\prime\prime}$}

\slugcomment{To appear in The Astrophysical Journal. Draft Oct. 21}

\shorttitle{IRS Observations of Debris Disks}
\shortauthors{Beichman etal.}

\begin{document}

\title{IRS Spectra of Solar-Type Stars: \break A Search for Asteroid Belt Analogs}

\author{C. A. Beichman$^{1,2}$, A. Tanner$^{2}$, G. Bryden$^{2}$, K. R. Stapelfeldt$^{2}$, M. W. Werner$^{2}$, G. H. Rieke$^{3}$, D. E. Trilling$^{3}$, S. Lawler$^{1}$, T. N. Gautier$^{2}$}

\affil{1) Michelson Science Center, California Institute of Technology,
M/S 100-22, Pasadena, CA, 91125}
\affil{2) Jet Propulsion Laboratory, 4800 Oak Grove Dr, Pasadena, CA 91109} 
\affil{3) Steward Observatory, University of Arizona, 933 North Cherry Ave, Tucson, AZ 85721}
\email{chas@pop.jpl.nasa.gov}

\begin{abstract}

We report the results of a spectroscopic search for debris disks surrounding 41 nearby solar type stars, including 8 planet-bearing stars, using the {\it Spitzer Space Telescope}. With accurate relative photometry using the  Infrared Spectrometer (IRS) between 7-34 $\micron$ we are able to look for excesses as small as $\sim$2\% of photospheric levels with particular sensitivity to weak spectral features. For stars with no excess, the $3\sigma$ upper limit in a band at 30-34 $\mu$m corresponds to $\sim$ 75 times the brightness of our zodiacal dust cloud. Comparable limits at 8.5-13 $\mu$m correspond  to $\sim$ 1,400 times the brightness of our zodiacal dust cloud. These limits correspond to material located within  the $<$1 to $\sim$5 AU region that, in our solar system, originates from debris associated with the asteroid belt.  We find excess emission longward of $\sim$25 $\mu$m from five stars of which  four also show excess emission at 70 $\mu$m. This emitting dust must be located around 5-10 AU. One star has 70 micron emission but no IRS excess.  In this case, the emitting region must begin outside 10 AU; this star has a known radial velocity planet. Only two stars of the five show emission shortward of 25 $\micron$ where spectral features reveal the presence of a population of small, hot dust grains emitting in the 7-20 $\mu$m band. The data presented here strengthen the results of previous studies to  show that excesses at 25 $\micron$ and shorter are rare: only 1 star out of 40 stars older than 1 Gyr or $\sim 2.5$\% shows an excess.  Asteroid belts 10-30 times more massive than our own appear are rare among mature, solar-type stars.

\end{abstract}


\section{Introduction}

The infrared excess associated with main sequence stars \citep{aumann1984} is produced by dust orbiting at distances ranging from less than 1 to more than 100 AU. In discussing these debris disks we often use the analogy of our Solar System, where dust is generated by collisions between larger bodies in the asteroid and  Kuiper belts, as well as outgassing comets. As in the Solar System, we distinguish between: a) hot (200-500 K) debris located relatively close to the star ($<$1 AU out to $\sim$ 5 AU) and arising in a region dominated by refractory material, e.g. debris associated with an asteroid belt \citep{dermott02};  and b) cooler debris ($<30-200$ K) located beyond 5 AU that may be associated with more volatile, cometary material, i.e. the analog of our Kuiper Belt. Observed dust luminosities range from $\ld \simeq 10^{-5}$ to greater than $10^{-3}$ for main-sequence stars, compared to $\ld \simeq 10^{-7}-10^{-6}$ inferred for our own Kuiper belt  \citep{stern1996,  backman1995} and $\sim 10^{-7}$ measured for our asteroid belt \citep{bp1993, dermott02}. 

The FGK survey of solar-type stars is a \spit\ GTO program designed to search for excesses around a broad sample of 148 nearby, F5-K5 main-sequence field stars,   sampling wavelengths from 7-34 \um\ with the Infrared Spectrograph (IRS; Houck et al. 2004) and 24 and \70um with  the Multiband Imaging Photometer for Spitzer (MIPS; Rieke et al. 2004). In the context of our  paradigm,  IRS is well suited to searching for and characterizing hot dust in an asteroid belt analogue while MIPS is better suited to studying cooler material in a Kuiper Belt analogue. The sample consists of two components --- stars with planets and those without. The MIPS observations of planet-bearing stars are discussed in Beichman et al. (2005a) and of non planet-bearing stars in Bryden et al. (2006).  This paper presents the results of the IRS survey of a sub-sample of 39 stars, 7 with planets, and the remainder without planets based on current knowledge from radial velocity programs. In addition, the paper includes IRS follow-up observations of 2 stars (one of which is planet-bearing) that were not originally scheduled for spectroscopy,  but which were found to have a 70 $\mu$m excess. Results for one star with a notable IRS spectrum, HD 69830, have been presented elsewhere \citep{beichman2005b} but will be summarized below in the context of the complete survey.

The IRS portion of the program was designed with four goals in mind: 1) to extend the wavelength coverage beyond that possible with MIPS to look for dust  emitting either shortward or longward of MIPS's $24\, \mu$m band; 2) to take advantage of IRS's broad wavelength coverage to reduce systematic errors and thus, possibly, identify smaller excesses than previously  possible; 3) to constrain the spatial distribution of dust grains by assessing the range of temperatures present in a given debris disk; and 4) to search for mineralogical features in the spectra of any excesses. In this paper we describe the sample of stars ($\S2$);  present our calibration procedure and give the reduced spectra for each star ($\S3$ and Appendix-1); discuss detections of or limits to IR excesses ($\S4$); present models of the excesses in the 4 stars with IRS and MIPS 70 um excesses and discuss the nature of their debris disks; discuss overall results and conclusions ($\S5$ and $\S6$).

\section{The Sample}

The details of the sample are presented in Bryden et al. (2006). In brief, the survey consists of solar-type stars with main sequence spectral types ranging from F5 to K5, most of which are within roughly 25 parsecs. Some stars at larger distances were included because they were known to possess planets via radial velocity studies \citep{butler1999, marcy2002}. A sub-sample of 39
stars covering the full range of spectral types was selected for IRS
observations. Table~\ref{basictable} describes 41 stars including follow-up observations of two stars earlier found by MIPS to have excesses (Table~\ref{mipstable}) \citep{beichman2005a, bryden2006}. Of the 39  original survey stars, seven are known to have planets from radial velocity studies with one additional planet-bearing star (HD 128311) selected for follow-up due to a possible MIPS excess. We also report MIPS results obtained after the observations reported in \citep{beichman2005a, bryden2006} for 9 stars in the spectroscopic sample (Table~\ref{newmipstable}); three of the stars observed with IRS have not yet been observed with MIPS due to the as yet incomplete scheduling of the GTO program. Of the 9 new stars observed with MIPS, none has a 24 or 70 $\micron$ excess.

\section{Observations and Data Reduction}

\subsection{IRS Spectra}

We observed each star with the three longest wavelength modules of the IRS: ShortLo-1 (SL1, 7.5-14.0 $\micron$), LongLo-2 (LL2, 14.0-20.5 $\micron$) and LongLo-1 (LL1, 20-34 $\micron$). The basic observing sequence and associated data reduction have been described in Beichman et al. (2005b) and are reviewed in the Appendix. In summary, we have used the fact that the majority of the sample shows no excess at MIPS wavelengths (24 or 70 $\micron$) or in an initial examination of the IRS data to derive a ''super-flat'' to improve the relative and absolute calibration of all the IRS spectra. After calibration with respect to photospheric models fitted to shorter wavelength ($\lambda\, <5\micron$) photometry, the agreement between the Spitzer/IRS data with IRAS and Spitzer/MIPS photometry is excellent:  IRS/IRAS(12$\micron$)= 0.96$\pm$0.01; IRS/IRAS(25$\micron$)=0.93$\pm$0.01; and IRS/MIPS (24$\micron$)= 0.99$\pm$0.03.

To look for IR excesses we first examined differences between the flattened,  calibrated spectra and photospheric models appropriate to each star (see Appendix-1 and Bryden et al. (2006)). For convenience, we define two "photometric bands" useful for isolating either the silicate features (8.5-13 $\micron$) or a long wavelength excess (30-34 $\micron$). For all 41 stars in the sample, the average level of the deviation from an extrapolation of the photospheric emission,  $(F_\nu(Observed)-F_\nu(Photosphere))/F_\nu(Photosphere)$,  has values of $-0.007\pm0.067$ in the 8.5-13 $\micron$ band and $-0.006\pm0.066$ in the 30-34 $\micron$ band. Only HD69830 \citep{beichman2005b} stands out as having a significant  excess from this first look at the data. 

However, a closer examination of the individual spectra showed that in most cases the data for an individual IRS module differed from the extrapolated photospheric emission in a systematic way that could be attributed to a simple scale or offset error. Without worrying about  the source of the scale error (due to either the photospheric extrapolation or residual calibration errors in the IRS spectra), we derived a new calibration factor for each star and for each IRS module using  the first 10 points of each module, effectively pinning the IRS spectrum to the photosphere at the short wavelength end of each of the three IRS modules: SL1 from  7.5 to 8.0 $\micron$; LL2 from  14.0 to 19.8 $\micron$; and LL1 from 19.5 to 21 $\micron$. Values of this calibration factor deviated from unity on a star-by-star basis by less than 10\%. This technique produced much smaller residuals showing no significant deviations from zero over the entire IRS wavelength range (Figures-1a and b) for the vast majority of the sample. The dispersion in the deviation from a smooth photosphere was reduced from $\sim7$\% to 1.3\% (8.5-13 $\micron$) and 1.7\% (30-34 $\micron$).\footnote{We also examined a star-by-star normalization factor derived using the entire module, not just the first 10 points. This choice resulted in a similar reduction in dispersion, but had deleterious effects on the few sources identified with excesses rising to longer wavelengths.} A longer wavelength excess beyond the first ten data points was not affected by the choice of the number of  points used to set the normalization. Using this technique, we found no deviation from a stellar photosphere for the majority of stars. However, as described below, we did find clear evidence of an excess rising up at wavelengths longward of  $\sim25 \,  \micron$ in five stars and a silicate feature appearing longward of 8 $\micron$ in two stars.

\begin{figure}[ht]
\epsscale{0.95} \plotone{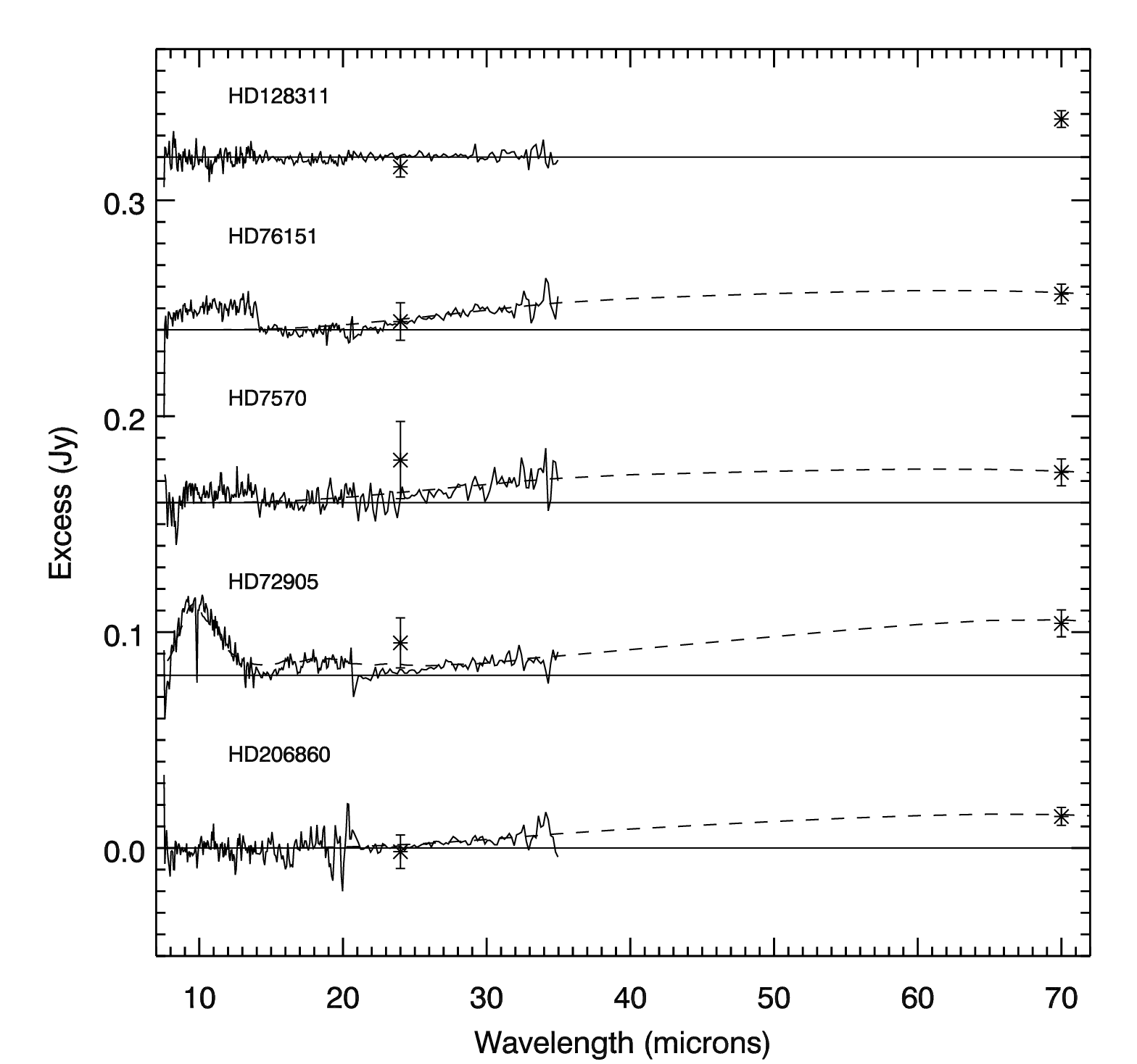} \caption{IRS spectra of those
stars with excesses (solid line) along with their best fitting
models (dashed line). The offset in SL-1 spectrum of HD 76151 is not statistically significant in terms of its fractional excess(1.4$\sigma$). The spectrum of HD 69830,  with its prominent IRS excess, is presented in Beichman et al. 2005b. \label{spec10}}
\end{figure}

Unlike the other stars in the survey, HD101501 showed anomalous  structure at the few percent level in its spectrum which could not be removed using the  normalizations described above. The module-wide deviations from  flatness are suggestive of uncorrected flat-field errors, not astrophysical effects. We put an upper limit of 5\% of the photospheric emission on any excess for this star. One or two stars, e.g. HD76151 (Figure~\ref{spec10}), showed small residual offsets of a few mJy between the spectrum and the photosphere in the 8-14 $\micron$ band where the star is brightest. As these ''features" are at a level less than the $\sim$1.5\% dispersion in the excesses averaged over the entire sample, we do not regard them to be significant.

\begin{figure}[ht]
\epsscale{0.95} \plotone{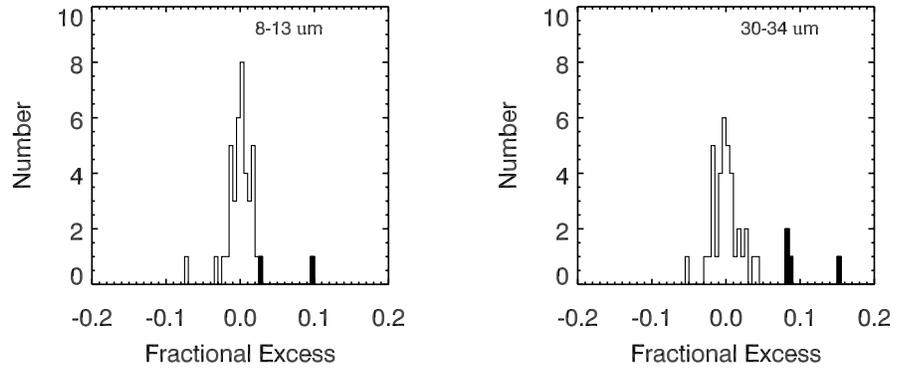}

\caption{Histograms showing the distribution of fractional excess  {\it ((Observed-Photosphere)/Photosphere})  estimated within pass bands at 8.5-13 (left) and 30-34 $\micron$ (right). The point for HD 69830 is off-scale to the right at 30-34 $\micron$ with a relative excess of 0.5 \label{histplot}}
\end{figure}

After flattening and normalizing the IRS spectra as described above, we estimated the fractional excess  $(F_\nu(Observed)- F_\nu(Photosphere))/ F_\nu(Photosphere)$.  Figure~\ref{histplot} gives histograms of the fractional  excess estimated in 8.5-13 and 30-34 $\micron$ "pass bands". In assessing the significance of an excess we looked at the internal uncertainty in the flux density measurement of a given star  and the fractional excess relative to the $\sim1.5$\% dispersion of the entire sample (Table~\ref{filtersummary}). Typically, the uncertainty in the fractional excess relative to the population as a whole is more important in assessing the reality of a spectral feature. There are a number of stars which appear to have a significant excess when looking only at the internal uncertainties in flux density, but which are not so impressive when compared to the dispersion in the overall population. If we require any excess to be more significant than 3 times the population averaged dispersion (1.3\% 8-13 $\micron$ and 1.7\% 30-34 $\micron$, Figure~\ref{histplot}), then based on the data presented in Table~\ref{filtersummary} we can claim statistically significant 30-34 $\micron$ excesses for five stars (HD 7570, HD 69830, HD 72905, HD 76151,  HD 206860). One star, HD 69830, shows a highly significant silicate feature starting around  8 $\micron$. Four of these sources, but not HD 69830,  also show a 70 $\micron$ excess. In the case of HD 72905 which also shows a 70 $\micron$ excess, its 8-13 $\micron$ flux density excess is highly significant relative to internal uncertainties (12$\sigma$), its fractional excess (2.8\%) is modestly significant (2.2$\sigma$) relative to the dispersion in the population, and the spectrum (discussed below) closely matches that expected for silicate emission. Thus, we regard the 8-13 $\micron$ excess of HD 72905 as real.

Figure~\ref{three}a-f shows the IRS spectra for six sources (HD 128311, HD 216437, HD 7570, HD 72905, HD 76151,  HD 206860) plotted in three different ways: a) their calibrated spectrum; b) their excess relative to photosphere after normalization with respect to the first 10 points of each module, and c) the fractional amount of the excess relative to the photosphere after normalization. The dashed lines in the  bottom panel show an estimate of the 2$\sigma$  dispersion in the deviations from the photospheric  models. Fractional deviations between these lines should be regarded with skepticism. The first two  stars (HD 128311, HD 216437) show no apparent excess and are representative of the bulk of the sample. The last four stars (HD 7570, HD 72905, HD 76151,  HD 206860, plus HD69830 shown in Beichman et al. 2005b) have excesses based on statistical criteria defined above. 

\subsection{New MIPS Data}

MIPS observations for  nine stars not previously reported \break \citep{beichman2005a, bryden2006} are summarized in Table~\ref{newmipstable}. The 24 and 70 \micron\ photometry were derived using processing steps detailed in Bryden et al. (2006) and is based on the DAT software developed by the MIPS instrument team \citep{gordon2005}. For consistency, we use the same analysis tools and calibration numbers as were adopted by Beichman et al. (2005a) and Bryden et al (2006).

At \24um, we carried out aperture photometry on reduced images using an
aperture of 6 camera pixels (1 pixel=2.5\arcsec \ at \24um)  in radius, a background annulus from 12 to 17 pixels, and an aperture correction of 1.15.  The flux level is calibrated at 1.047 $\mu$Jy/arcsec$^2$/(DN/s), with a default color correction of 1.00 appropriate for sources warmer than 4000 K at a weighted-average wavelength of 23.68 \um.  

At \70um we used images processed beyond the standard DAT software in
order to correct for time-dependent transients, corrections which can significantly improve the sensitivity of the measurements
\citep{gordon2005}. Because the accuracy of the \70um data is limited by background noise, we used a relatively  small aperture  of just 1.5 pixels (1.5$\times$9\arcsec) in radius and a 4 to 8 pixel radius sky annulus. This aperture size requires a relatively high aperture correction of 1.79.  The flux level is calibrated at 15,800 $\mu$Jy/arcsec$^2$/MIPS\_70\_unit,  again with a default color correction of 1.00.  

The 24 and 70 $\mu$m data are presented in Table ~\ref{newmipstable} which also includes a comparison with the predicted level of photospheric emission based on models fitted to 2MASS and shorter wavelength data \citep{bryden2006}. None of the new objects shows an excess at 24 $\micron$. However, since accurate 2MASS data are not available for most of these bright ($K_s<6$ mag) stars due to saturation effects, a significant part of the dispersion for the stars studied herein is probably due to the uncertainties in the photospheric extrapolation.  We define the parameter $\chi70= (F_\nu(Observed)-F_\nu(Photosphere))/\sigma$ at 70 $\micron$ where $\sigma$ is a combination of internal and calibration uncertainties as discussed in detail in Bryden et al. (2006). Using $\chi70$ as a criterion, none of the new objects show a significant 70 $\micron$ excess.

The 30-34 $\micron$ excess toward  the five stars described above (HD 7570, HD 69830, HD 72905, HD 76151,  HD 206860) have been carefully compared with MIPS observations at 24 $\micron$ \citep{bryden2006}. A sample of 87 stars in the FGK survey  selected for having no 70 $\micron$ excess (MIPS-70/Photosphere$<$1.5) has a 2-$\sigma$ clipped average of the MIPS-24/Photosphere ratio=0.978$\pm$0.001 with a dispersion of 0.056 \citep{bryden2006}. HD 69830 is clearly detected as a strong MIPS-24 excess with a MIPS-24/Photosphere $\sim 1.5$. HD 76151 and HD 206860 have MIPS-24/Photosphere ratios of 1.00 and 0.96 and show no evidence for 24 $\micron$ excess. HD 72905 and HD 7570 represent intermediate cases, however, with MIPS-24/Photosphere ratios of 1.064 and 1.053. While the nominal dispersion for the 24 $\micron$ data would suggest that these MIPS-24 excesses are not significant ($<2\sigma$), further analysis of the MIPS data (Kate Su, 2005, private communication) suggests that the combined photospheric and MIPS uncertainties may be closer to 2\% in which case these modest excesses may turn out to be real. In a similar vein, Bryden et al. (2006) have identified a weak general trend ($1.5\sim2\sigma$) for stars with 70 $\micron$ excesses to have slightly  higher MIPS-24/Photosphere ratios than for stars without a longer wavelength excesses. If the weak MIPS excesses hinted at here can be made more significant with careful calibration and validation using IRS, then it may be possible to look for weak, warm excesses in a much larger sample of objects than would be possible with IRS alone.

\section{Results}

\subsection{Statistics of detections}

We detected significant IRS excess emission in five sources including HD
69830 (Beichman et al. 2005b). Of these five stars, all except HD69830 also have a MIPS 70 $\micron$ excess. Two stars, HD 128311 and HD 206860, were included explicitly for IRS follow-up observations because they had MIPS excess emission. HD 128311 shows no IRS excess\footnote{The  70 $\micron$ detection of HD 128311 \citep{beichman2005a} is marginal and awaits confirmation in a pending MIPS observation.} while HD 206860 does show an excess at 30-34 $\micron$. Out of the 39 stars in the initial survey, we have identified 30 $\micron$ IRS excesses around four stars (HD69830, HD72905, HD7570, HD76151), or 11$\pm$5\%, which is consistent with the fraction of stars with excesses found at 70 $\micron$ \citep{bryden2006}.  Including the two follow-up stars (HD206860 and HD 128311) brings the detection rate to 5 of 41 stars or 12$\pm$5\% at 30$\micron$. None of the eight planet-bearing stars have 30 $\micron$ excesses, a smaller fraction than seen with MIPS 70 $\micron$ excesses,  $<12\%$ vs. 24$\pm$10\% \citep{beichman2005a}. The result is suggestive of an absence of warm dust at levels detectable by IRS,  but is not inconsistent with the expected fraction given the small number statistics. 

If we apply a similar analysis to the 8-14 $\micron$ portion of the spectrum, the corresponding statistics give detections of just two objects (HD69830 and HD72905) out of the 39   stars (with and without planets) in the initial survey, or 5.1$\pm$3.6\%, and zero of eight planet-bearing stars. Removing HD 72905 from the overall statistics because it is quite young, 300 Myr \citep{spangler2001} based on its membership in the Ursa Major moving group, reduces the statistics to just 1 short wavelength excess out of 38 mature stars ($>1$ Gyr as discussed in Bryden et al 2006) or $<$2.5\% in the initial survey. These data confirm the rarity of short wavelength excesses compared with ones at longer wavelengths. This result was first noted on the basis of IRAS and ISO data \citep{aumann1991, mannings1998, fajardo2000, laureijs2002} and reconfirmed with MIPS  data \citep{beichman2005a, bryden2006}, but is extended here to  lower levels of emission, i.e. 2\% relative excess for IRS vs. 6-10\% for IRAS and  ISO. In the next section we convert these observational limits into a limit on the fractional luminosity of dust in these systems, $\ld$.

\subsection{Limits on the Fractional Disk Luminosity, $\ld$}

A useful metric for the limits on dust surrounding these stars is  $L_d/L_*$,
which is related to the fractional flux limit of an excess relative to the Rayleigh Jeans tail of the star's photosphere (Bryden et al. 2006):

\begin{equation}
\frac{F_{{\rm dust}}}{F_{\star}} = \frac{L_{\rm dust}}{L_{\star}} \;
\frac{h\nu T_{\star}^3}{kT_{\rm dust}^4 (e^{h\nu/kT_{\rm dust}} -1)}
\end{equation}

\begin{figure}[ht]
\epsscale{0.8} \plotone{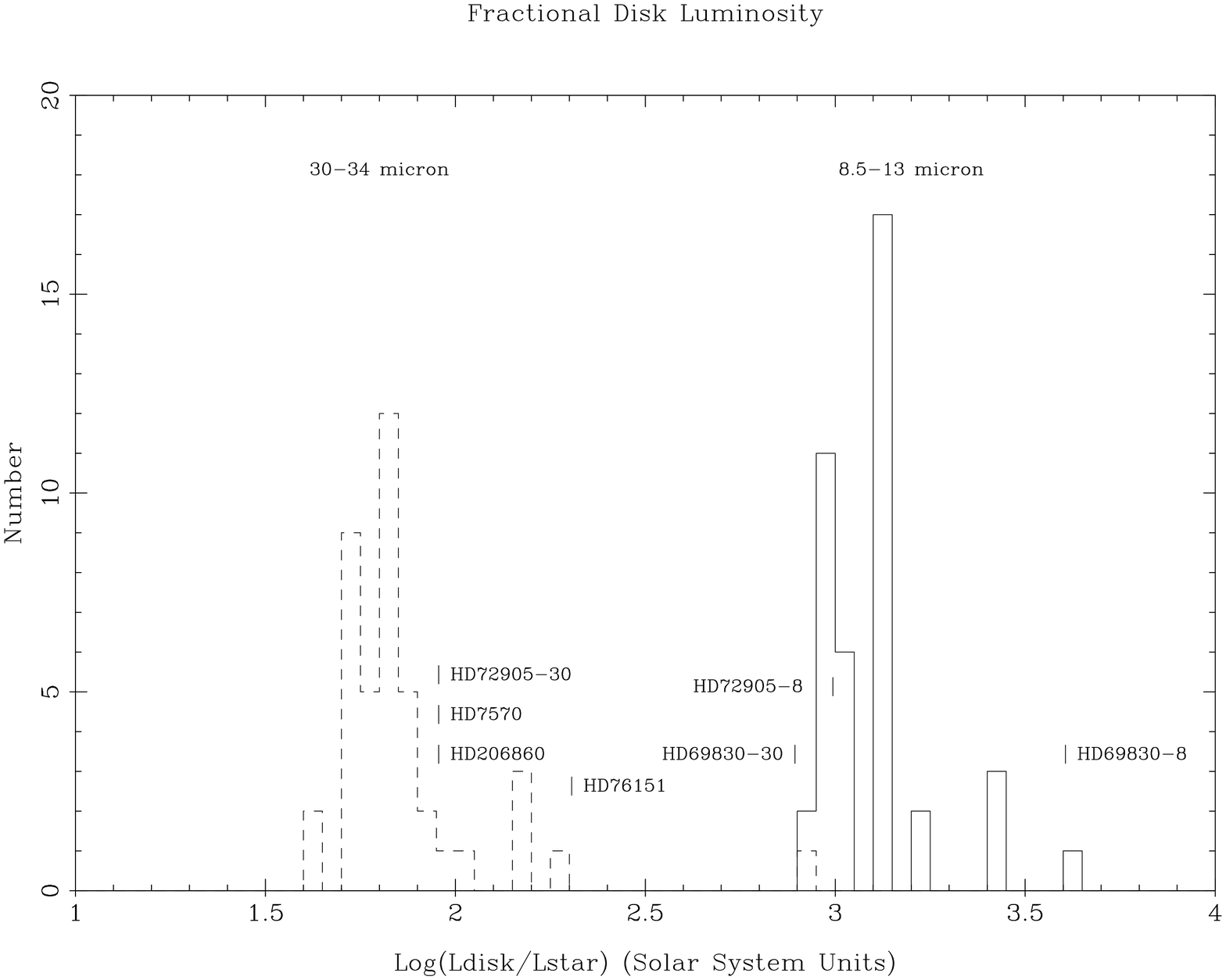} \caption{
Histograms showing the detections or (mostly) 3$\sigma$ upper limits on $\ld$ for dust emitting in  30-34 $\micron$ band (dashed lines) and the 8.5-13 $\micron$ band (solid lines) for the stars in this survey. These limits take into account the effective temperature, $T_{eff}$,  and fractional excess for each star as described in the text and are given in units of the solar system's approximate level of zodiacal emission, $\ld=10^{-7}$ \citep{bp1993}. The positions of the positive detections are indicated along with the star's name. An $-8$ or $-30$ is appended to the name to resolve any ambiguity as to which band is being shown. The remainder of the values are upper limits. \label{filtersummaryfig}}
\end{figure}

From which it is easy to show that

\begin{equation}
{ {L_d} \over {L_*} }  = { {F_{\rm dust}} \over {F_*} } { { e^{x_d}-1} \over {x_d} }  \left (  {{T_d} \over {T_*}}\right ) ^3
\end{equation}

\noindent where $F_{\rm dust}=F_\nu(Observed)-F_\nu(photosphere)$. At the peak of the black body curve  $x_d\equiv h\nu/kT_d$ has a constant value of 3.91, corresponding to $T_d=367$ K at 10 $\micron$. At this wavelength $L_d/L_* = 0.00345 \left (  {{5600 K} \over {T_*}} \right ) ^3  F_d/F_*$. In Table~\ref{filtersummary} and Figure~\ref{filtersummaryfig}, we evaluate  $\ld$ for each star using the appropriate effective temperature and luminosity (from SIMBAD, Table~\ref{basictable}) and its measured fractional excess, $F_{\rm dust}/F_*$, or in the case of an upper limit,  $3\sigma_{pop}$ where $\sigma_{pop}$ is the dispersion in fractional excess averaged over the whole sample. We calculate these dust limits in the 8.5-13 $\mu$m and the 30-34 $\mu$m portions of the IRS spectrum assuming $T_d=367$K and $T_d=115$K at 10 and 32 $\micron$, respectively. This definition $\ld$ assumes that the emitting material is all located at the location where the peak of the $T_d$ blackbody matches the wavelength of observation. More emitting dust, and higher values of $\ld$, would be required for material located substantially interior or exterior to this point. 

The 3$\sigma$ limits on $\ld$ at 8.5-13 $\micron$ and 30-34 $\micron$ have 2$\sigma$ clipped average values of $\ld=14\pm 6\times 10^{-5}$ and $0.74\pm 0.31 \times 10^{-5}$ respectively (Table~\ref{filtersummary}). In comparison with our solar system, $\ld \sim$ 10$^{-7}$ \citep{bp1993,dermott02}, the IRS results set limits (3$\sigma$) on warm (360 K) dust peaking at 10 $\micron$ of $\sim$1,400 times the level of emission in our solar system. For cooler dust ($\sim$ 120 K) peaking at 30-34 $\micron$, the 3$\sigma$ limit corresponds to $\sim$74 times  the nominal $\ld$ of our zodiacal cloud. 

For objects with detections in the IRS bands, we can determine $\ld$ explicitly by integrating over the data between 10-34 and extrapolating using the models discussed below out to 70 $\micron$. These values are given in Table~\ref{mipstable} and are compared with the 70 $\micron$-only estimates obtained using equation (1). Including the IRS spectra increases the estimated $\ld$ by about a factor of two relative to the 70 $\micron$-only estimate. For the nine stars with new  upper limits to an excess at 70 $\mu$m (Table~\ref{newmipstable}), we also report limits on $\ld$. 

\section{Discussion}

\subsection{Models of Disks around Solar Type Stars}

We fitted the IRS spectra and MIPS 70 $\micron$ photometry using a simple model of optically thin dust within one or two flat dust annuli centered around the star. As a guide to subsequent modeling, we first modeled the emission using blackbody grains at a single temperature in equilibrium with the parent star. From the ratio between the 30-34 $\micron$ and the 70  $\micron$ flux densities for each star we found (Table~\ref{modtable}) grain temperatures in the range 75-100 K with equilibrium distances of 8-14 AU for the four stars. Even though the uncertainties in the flux density ratio are  quite large ($\sim 25\%$), the temperature and location of the grains are reasonably well defined, $\pm$1 AU and $\pm$15 K, given the assumption of blackbody grains.

We calculated the power-law temperature profiles, $T(r)=T_0 (L/L_\odot)^\alpha (r/r_0)^\beta$, for these grains  as a function of stellar luminosity and distance from the central star by solving for the grain temperature at which the integrals over the absorbed stellar radiation and emitted thermal radiation were equal. The equilibrium temperature and power-law variation of temperature with distance and stellar luminosity depend on the assumed grain size and composition \citep{bp1993}. As described below, we use 10 and 0.25 $\mu$m silicate grains \citep{draine1984, weingartner2001} for which the we obtained the following numerical relationships for large and small grains, respectively: $T(r)=255 \, K (L/L_\odot)^{0.26} (r/AU)^{-0.49}$ and $T(r)=362\, K (L/L_\odot)^{0.29} (r/AU)^{-0.41}$. These calculated coefficients and power-law constants closely follow analytical results \citep{bp1993}.

We then calculated the dust excess by integrating over the surface brightness of a disk between $R_1$ and $R_2$, with $F_\nu(\lambda)={ {2\pi} \over {D^2} } \int \tau_0(\lambda) (r/r_0)^{-p} B_\nu(T(r)) r dr $.  In calculating the wavelength-dependent dust optical depth, $\tau(\lambda)$, we adopted grain emissivities $Q_{abs}$ appropriate to small (0.25 $\micron$) or large (10 $\micron$) grains \citep{draine1984, weingartner2001}. The integrated surface density in the  wedge-shaped disk expected for grains dominated by Poynting-Robertson drag is roughly uniform with radius, $p=0$ \break \citep{burns1979, buitrago1985, backman2004}. We examined a number of other cases with $0<p<1$ that would reflect different dust dynamics, but did not find results that were substantially different from those for $p=0$.  Results of the model fitting are shown in Figure~\ref{spec10} and Table~\ref{modtable} and are discussed below. 

Two other Spitzer studies have recently been published with similar results to those found here. Hines et al (2006) identified a much younger ($\sim 30$ Myr) star, HD 120839, with an IRS spectrum similar to that of HD69830. Reporting on early results from the FEPS survey \citep{meyer2004}, Kim et al (2005) found a number of stars with 70 $\mu$m excesses with, in some cases, an accompanying excess in the IRS bands. The model results concerning the amount of dust and its orbital location are similar to those presented here.

\subsubsection{Single Ring Models for HD 76151, HD7570, HD 206860, and HD128311}

For HD 7570, HD 206860 and HD 76151, which show no statistically significant,  short wavelength excess, we used only a single population of large, 10 $\mu$m, amorphous, silicate grains \citep{draine1984, weingartner2001}, fitting emission from a single annulus to 89 data points longward of 21 $\micron$ (just longward of the last point used for the flux normalization of the LL-1 IRS module) and including 70 $\micron$.  By varying $\tau_0$, $R_1$ and $R_2$ we were able to minimize the reduced $\chi^2$ to values between 0.5-0.8 with $77-3=74$ degrees of freedom.  

The models for HD 76151 and HD 7570 require grains with a narrow spread of temperatures around 100 K and located in the range of 7-10 AU. Assuming from 10 $\micron$ grains, the predicted 850 $\mu$m flux density from the HD 7570 disk is well below the 1  mJy limit from SCUBA (J. Greaves, private communication, 2005). The predicted 850 $\micron$ flux density from a disk containing larger ($\sim 100\, \micron$) grains this system would be approximately 1 mJy, just at the SCUBA limit. We cannot distinguish between the two grain sizes using only  Spitzer data. HD 206860 has cooler grains (75 K) than the other stars with material located around 14 AU.

Mass estimates are notoriously tricky to derive given uncertainties in grain sizes, but these three systems suggest dust masses around 1-2.6 $\times 10^{-6} M_\oplus$ (Table~\ref{modtable}) for a silicate grain density of 3.3 gm cm$^{-3}$. Extrapolating this estimate using the $-3.5$ index power law appropriate for a distribution of sizes from a collisional cascade \citep{dohnanyi1969} up to a maximum size, $R_{max}$, of 10 km yields total mass estimates of 0.024-0.080$\sqrt{R_{max}/10 {\rm km}} \, M_\oplus$. Submillimeter observations of all these sources would further constrain the dust size and distribution and thus the total mass of the emitting material.

Since the IRS data place only a limit on the emission shortward of 34 $\micron$ for the planet-bearing star HD 128311, the single grain model can be used to show that the inner edge of the disk seen at 70 $\micron$ must start beyond  $\sim$ 15 AU corresponding to material cooler than $\sim 50$K. 

\subsubsection{Two Ring Model for HD72905}

Since HD 72905 shows evidence for emission from small silicate grains starting around 8  $\micron$, we separately examined three models: 1) small grains over a broad range of distances; 2) small grains in two annuli, one close to the star and one further away; 3) small grains in a close-in annulus and large grains in a more distant annulus. For the small grains we used 0.25 $\micron$ grains composed of either amorphous silicates or crystalline forsterite \citep{malfait1998, jager1998}.  Unlike the dramatic case of HD 69830, small crystalline silicate grains provide a poor fit to the 8-14 $\micron$ data for this star (Figure~\ref{spec10}). Amorphous silicate grains have a broad emission profile particularly on the short-wavelength side which appears in the HD 72905 spectrum but not in HD 69830's. A model providing a good fit to the data incorporates amorphous silicate grains located in an annulus extending from the grain sublimation distance (0.03 AU corresponding to 1500 K) out to 0.43 AU where the characteristic temperature is 500 K. The spectrum suggests an upper limit of no more than 10\% of crystalline forsterite grains (by number).

However, this small dust population cannot extend out continuously much beyond 0.4 AU. Even with a density falloff proportional to $r^{-0.4}$ to $r^{-1}$, the presence of any dust between 1 and 10 AU produces far more radiation in the 20-70 $\micron$ region than is observed in the combined IRS+MIPS data. Thus we reject model \#1 above. Rather, the data longward of 14 $\micron$ suggest a second population of cooler material located beyond 10 AU. We examined models with either small grains (0.25 $\micron$, model \#2) or large grains (10$\micron$, model \#3) emitting at temperatures less than 90 K. Both models \#2 and \#3 give reasonable fits to the observations with reduced $\chi^2$'s $\sim0.8$. As Table~\ref{modtable} indicates, a much greater quantity of small grains emitting at cooler temperatures is required to fit the data compared to a model using larger grains. This difference is due to the greatly reduced emissivity of small 0.25$\micron$ vs. large 10$\micron$ grains, e.g.  at 70 $\micron$, $Q_{abs}$=0.008 vs. 1.84 \citep{draine1984, weingartner2001}. 

   A testable difference between the large and small grain models is the size of the emitting region. For small grains, the emitting annulus extends between 109 and 310 AU, corresponding to a diameter of 40\arcsec\ at the 14.1 pc distance to the star. By contrast, large grains cool more efficiently and are located closer to the star, $<16$ AU, for a predicted diameter of only 2\arcsec. Since there is no evidence that the 70 $\micron$ image of HD 72905 is extended, we favor the large grain model. The total amount of material in the HD 72905 disk is about a factor of two greater than in the other disks studied here.

\subsection{Nature of Disks}

\subsubsection{Rarity of Emission from Warm Dust}

The vast majority of the stars studied here, planet-bearing or not, have no emission interior to $\sim$ 10 AU at the levels to which IRS is sensitive, 75-1,400 times that of our zodiacal cloud. Since this result applies equally to stars with and without planets, there must be some general explanation for the absence of warm material around mature stars ($>1$ Gyr). A long term decline in the amount of debris expected on theoretical grounds \citep{dominik2003} and demonstrated with the envelope of the distribution of 24 $\micron$ excesses seen in A stars, implying an excess $\propto { {150 Myr} \over {t_{Myr}} }$  \citep{rieke2005}, is consistent with the results of this survey. Modifying Eqns. (35)-(40) of Dominik and Decin (2003) for parameters appropriate to an asteroid belt located between 1-5 AU leads to the conclusion that after one billion years, $\ld$ in the IRS wavelength range should be $\leq 10^{-6}$ and thus undetectable with IRS over a broad range of initial conditions. 

Three of the disks (HD 7570, HD 206860, HD76151) have an inner edge around  $\sim10$ AU and may be only  a few AU wide (Table~\ref{modtable}). The cutoff at wavelengths shorter than $\sim$25 $\micron$ corresponds to temperatures $<$ 75-100 K and location $\geq 7$ AU. Other stars observed in this or other Spitzer programs with 70 $\micron$ excess that do NOT show an inner emission cutoff at 25 $\micron$ seem to have some extenuating circumstance: Vega with evidence for a recent collision \citep{su2005}; $\beta$ Pic which is extremely young; Fomalhaut \citep{stapelfeldt2004}, two young stars HD 72905 and HD 12039 \citep{hines2006} and HD 69830 \citep{beichman2005b} with a sharply bounded, interior dust belt originating, perhaps, in an active asteroid belt.  If the observational trend for a general lack of material interior to $\sim$10 AU continues as more Spitzer data accumulates, then it may be that this inner boundary to the "Kuiper Belt" is related to the ice-sublimation distance $\geq 3L_\star^{0.5}$ AU where significant amounts of material could have accumulated in the protoplanetary disk \citep{thommes2003}.

The amount of warm material around HD 72905 with $\ld\sim 18\times 10^{-5}$ is well above values consistent with the simple models \citep{dominik2003} even given the relative youth of this star ($\sim$ 0.3 Gyr). The excess may be attributable to a recent collisional event between planetesimals as has been suggested by Kenyon and Bromley (2004) and Rieke et al. (2005) for A stars observed at 24 $\micron$. As discussed in Beichman  et al. (2005b), the disk of hot, small grains around a mature star, HD 69830, stands out as exceptional in this survey and may similarly be the result of a rare transient event in an unusually massive asteroid belt. In the case of steady-state emission (rather than transient events), the brightness of an excess will be proportional to the first or second power of the number of colliding planetesimals, depending on whether the ultimate loss mechanism for the smallest grains is blowout by radiation pressure or PR-drag, respectively \citep{dominik2003}. Since PR-drag is likely to be more effective than blowout for the low luminosity stars considered here \citep{sheret2004},  a 74-1,400-fold brighter zodiacal cloud than our own would imply the existence of (or a limit to) an asteroid belt $\sqrt{75}\sim \sqrt{1400}\sim 8-35$ times  more massive than our own, depending on its location. The results of this IRS survey suggest that asteroid belts of such a large mass  are rare around mature, solar type stars.

As pointed out by an anonymous referee, there are alternatives to the transient, collision hypothesis for the origin of the debris disks around mature stars. Stirring up a quiescent asteroid or Kuiper belt by planetary migration caused by either orbital interactions \citep{gomes2005, strom2005} or to the effects of a passing star (Kenyon \& Bromley 2004b) has been suggested as possible causes for analogues of  the period of Late Heavy Bombardment in our solar system. A strong infrared excess could be a signpost of a similarly exciting period  in other planetary systems. For the young stars HD 72905 (this survey) and  HD 120839 \citep{hines2006} this analogy is quite plausible. On the other hand, for the older stars in this survey, including the $>$2 Gyr-old HD 69830, it is not clear that planetary migration is likely so long after the formation of these planetary systems. Confirmation of this hypothesis will require a more complete census of planetary systems than is at present available.

\subsubsection{Lack of Warm Dust and the Presence of Planets}

\begin{figure}[ht]
\epsscale{0.95} \plotone{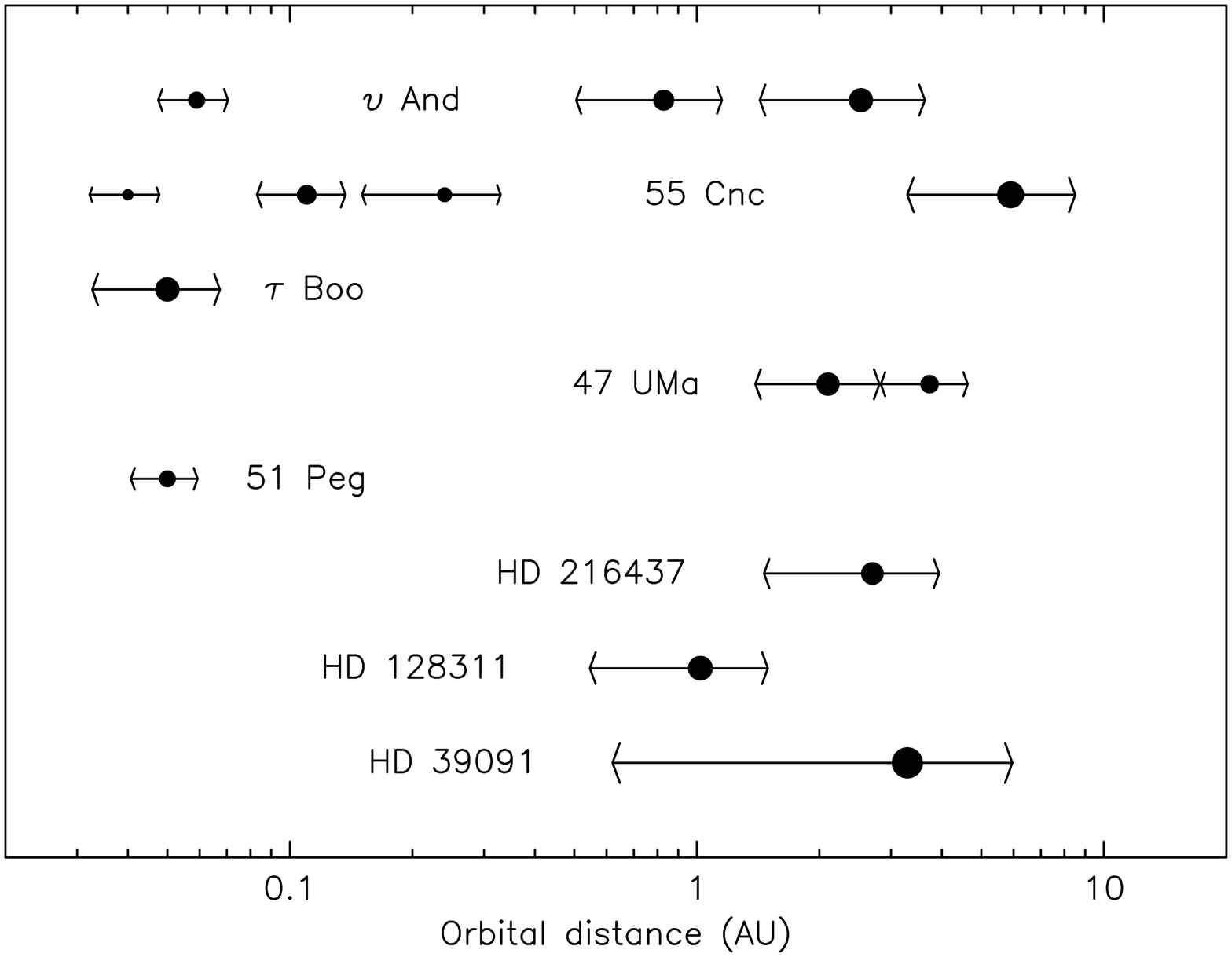}
 \caption{For each star with planets the minimum and maximum extents of the gravitational influence of that star's planets are plotted. The lack of emission in the IRS wavelength region is consistent with a dearth of material interior to about 5 AU. A partial explanation may be the influence of planets on dust in disks. \label{hillfig}}
\end{figure}

None of the planet-bearing stars show an IRS excess and only one, HD 128311, shows a MIPS excess. On the assumption that the gravitational influence of a planet at orbital semi-major axis, $a_p$, and eccentricity, $\epsilon_p$ is given by $r_{grav} \sim \sqrt{\epsilon_p^2 a_p^2 +12R_{Hill}^2}$  where the Hill Radius is defined as  $R_{Hill}=a_p\left ( M_p/(3 M_*) \right )^{1/3}$ \citep{bryden2000} with  $M_p$ and $M_*$ the masses of planet and star.  As shown in Figure~\ref{hillfig}, the region between $R_{Min}=a-r_{grav}$ and $R_{Max}=a+ r_{grav}$ could be swept clean of dust due to sweeping up action by the planet\footnote{Planet properties from Planet Encyclopedia at http://cfa-www.harvard.edu/planets}. In addition to more global effects leading to a general decline in dust content as discussed in the preceding section, gravitational effects of planets in relatively  distant orbits (HD 39091, 55 Cnc, 47 U Ma, HD 216437, and $\upsilon$ And) might clear out the material interior to $\sim$ 10 AU in the absence of a secondary source such as a massive asteroid belt. Stars with close-in planets ($\tau$ Boo and 51 Peg) may require another explanation (or an as yet undetected, more distant planet) for the lack of material in the 1-10 AU region.

\section{Conclusion}

We have used the IRS spectrometer on the Spitzer Space Telescope  to look for weak excesses around main sequence, solar type stars. After careful calibration, the IRS spectra allow us to achieve relatively low values of $\ld$ compared to previous studies. At 8-13 $\micron$ the IRS data provide an average 3$\sigma$ limit of $\ld =14 \pm 6 \times 10^{-5}$ or roughly  1,400 times the level of our solar system zodiacal cloud for material in the 1 AU region of the target stars.  Two stars, HD 72905 and HD 69830, have detectable emission at these wavelengths and in both cases there is evidence for emission by small dust grains. At 30-34 $\micron$, we reach a more  sensitive, limit on $\ld$ of $0.74 \pm 0.31 \times 10^{-5}$, or roughly  74 times the level of our solar system zodiacal cloud for material in the $5-10$ AU  region which may represent the inner edge of the Kuiper Belt. Four stars (HD 7570, HD 72905, HD 76151, HD 206860) showed evidence for long wavelength IRS emission that appears to link smoothly to a 70 $\micron$ MIPS excess. HD 69830 shows no evidence for 70 $\micron$ excess and thus for colder, more distant material.

The lack of IRS excess emission toward planet-bearing stars may be due to the small size of the sample observed to date, to the effects of planets removing dust from the orbital locations probed at IRS wavelengths, or to more general processes that remove dust from regions interior to a few AU \citep{dominik2003}. We suggest that asteroid belts located between 1-10 AU and as massive as 8-35 times our own asteroid belt are rare around mature, solar type stars.

\section{Acknowledgments}

Sergio Fajardo-Acosta  provided programs we used for accessing the
Kurucz models. This research  made use of the IRAS, 2MASS, and
Hipparcos Catalogs (1997), as well as the SIMBAD database and the VizieR
tool operated by CDS, Strasbourg, France. We gratefully acknowledge the comments of an anonymous referee whose careful reading of the manuscript led to a number of useful points of discussion and clarification.

The {\it Spitzer Space Telescope} is operated by the Jet Propulsion
Laboratory, California Institute of Technology, under NASA contract
1407.  Development of MIPS  was funded by NASA through the Jet
Propulsion Laboratory, subcontract 960785.  Some of the research
described in this publication was carried out at the Jet  Propulsion
Laboratory, California Institute of Technology, under a contract
with the National Aeronautics and Space Administration.

Finally, we remember with great sadness the efforts of NRC
postdoctoral fellow Elizabeth Holmes who worked intensively on this
project before her untimely death in March 2004.

\section{Appendix-1. Details of IRS Data reduction}

Standard IRS Staring mode observations were made with the Short Low Order 1 (SL1; 7-14~$\mu$m), Long Low Order 2 (LL2; 14-20~$\mu$m),  and Long Low Order 1 (LL1; 20-35~$\mu$m) modules. Each star was observed at two  positions along the slit, called Nod$_1$ and Nod$_2$. Data were processed by the Spitzer Science Center (SSC) to produce calibrated images of the spectrometer focal  plane. The data presented here were processed with Version 11.0 of SSC pipeline  (February 2005). The SSC SPICE software was used to extract spectra from the images. Because of the need for careful subtraction of the stellar continuum to detect a faint  excess, additional steps were taken in producing the final spectra. These are now described briefly.

The first few points at the beginning and end of the spectrum from each module were typically unreliable, as were a few bad pixels flagged in the SSC processing. These data were rejected. This effect was particularly noticeable at the long wavelength end of SL1 ($\lambda>14\, \mu$m). However, since the short wavelength end of LL2 overlaps SL1, there is no gap in the final spectrum.

Longward of $\sim$7 $\mu$m, the photospheres of solar type stars are smooth and do not differ greatly from a Rayleigh-Jeans  blackbody curve \citep{kurucz1992, castelli2003}. We used this fact to  improve the pixel-to-pixel calibration in the extracted SSC/SPICE  spectra. Of the 41 stars listed in Table~\ref{basictable}, 28 stars  showed no clear evidence for an excess at either 24 or 70 $\micron$ with MIPS or with the initial pass of IRS processing but did show consistent pixel-to-pixel deviations from a smooth photospheric model.

For these stars we formed ratios of the extracted IRS spectra to the Kurucz model appropriate for the effective temperature and metallicity of each star. From the average of these ratios at each wavelength, we created a $``$super-flat'' response curve for the SL1, LL2 and LL1 modules. Only those stars which did not show signs of an excess during the initial pass of the data were included in the super-flat. Shortward of 14 $\micron$,the SL1 super-flat typically has values in the range 0.97-1.03, or,equivalently, deviations from unity response of $\sim$3\%; a few pixels have values deviating from unity by 10\%. The values at each pixel of the SL1 super-flat have a dispersion around the pixel response (averaged over 28 stars showing no signs of excess in IRS or MIPS data) of $\sigma_{pop}\sim2\%$ and $\sigma_{mean}\sim0.4\%$. In the LL1 and LL2 modules, the typical super-flat values are in the range 0.98-1.02 with dispersions of $\sigma_{pop}\sim1-2\%$ and $\sigma_{mean}\sim0.2-0.3\%$. We divided the extracted spectra by the appropriate $``$super-flat'' to remove any residual calibration variations. The dispersion in the super-flat ($\sigma_{pop}$) is an indication of the limiting systematic noise in removing the photospheric contribution to the signal from these stars.

Although the spectrum of HD 101501 shows no excess at the $\sim$10\% level in either the MIPS or IRS data, its spectrum did not behave in a consistent manner when the various normalization steps were applied to look for fainter excesses. Differential pointing issues or other time variable effects could have produced a spectrum deviating from the mean for non-astrophysical reasons. The module-wide deviations from  flatness are suggestive of uncorrected flat-field errors, not astrophysical effects. As a result, this star was excluded from the generation of the super-flat.

The comparison between the SSC spectra and the photospheric models  fitted to shorter wavelength photometry \citep{bryden2006} shows absolute calibration gain factors of \break $F_\nu(SSC)/ F_\nu$(Photospheric Model)= $0.84\pm0.01$  (5\% dispersion from star to star) in the LL-1 and LL-2 modules and $1.12\pm0.014$ (7\% dispersion from star to star) in the SL-1 module. As discussed in $\S3.1$ the agreement between IRS data and IRAS or MIPS observations is excellent with deviations from unity of $4\pm$1\% at 12 $\micron$,  $7\pm$1\% at 24 $\micron$ (consistent with a known calibration issue with IRAS data at this wavelength, e.g. Cohen et al 1992),  and $1\pm$1\% at MIPS-24 $\micron$.

\clearpage
\begin{deluxetable}{ll|cccccc}
\tabletypesize{\scriptsize} 
\tablecaption{An IRS Survey of F, G, and K  Stars \label{basictable}}
\tablehead{Star$^1$&Name&Sp. Type&V(mag)$^2$&Dist(pc)&AOR&MIPS Excess$^3$&IRS Excess$^4$}
\startdata
HD 693 & GJ10 & F5V           &3.51&18.9&4008448&No. New& \\
HD 3795 & GJ799 & G3V          &3.86&28.6&4009216&N/A& \\
HD 4628 & & K2V &6.41&7.4&4009984&No& \\
HD 7570 & GJ55 & F8V          &4.07&15.1&4010240&70 $\micron$& Long\\
HD 9826$^5$  & $\upsilon$ And &   F8V  &4.09&1.35&4010496&No. New& \\
HD 10800 & GJ67.1 & G2V  &3.71&27.1&4010752&No& \\
HD 39091$^5$  & GJ9189&   G1V  &5.67&18.2&4015616&No& \\
HD 43834 & GJ231 & G6V &5.05&10.2&4015872&No& \\
HD 55575 & GJ1095 & G0V &4.42&16.9&4016128&No& \\
HD 58855 & GJ9234 & F6V &3.87&19.9&4016384&No& \\
HD 69830 & GJ302 & K0V &5.45&12.6&4016640&24 $\micron$& Short, Long\\
HD 72905 & GJ311 & G1.5V &4.93&13.8&4016896&70 $\micron$& Short, Long \\
HD 75732$^5$  & $\rho$ (55) Cnc A &   G8V&5.95&12.5&4017152&No& \\
HD 76151 & & G3V &5.74&11.3&4017408&70 $\micron$& Long\\
HD 84737 & GJ368 & G0.5Va &3.77&18.4&4017664&No& \\
HD 86728 & GJ376 & G3Va &4.53&14.9&4017920&No. New& \\
HD 95128$^5$  & 47 UMa &   G1V&5.1&14.1&4018432&No& \\
HD 101501 & GJ434 & G8Ve &5.43&9.5&4018944&No& \\
HD 120136$^5$  & $\tau$ Boo&   F6IV&4.5&15.6&4021760&No& \\
HD 133002 & GJ3876 & F9V &2.46&43.3&4022272&No& \\
HD 136064 & GJ580.2 & F8V &3.11&25.3&4022784&No& \\
HD 142373 & GJ602 & F9V &3.61&15.8&4023040&No& \\
HD 146233 & & G1V &4.56&15.4&4081920&No& \\
HD 154088 & GJ652 & G8IV-V &5.3&18.1&4029952&No. New& \\
HD 166620 & & K2V &6.14&11.1&4023808&No& \\
HD 168151 & & F5V &3.18&23.5&4024064&No& \\
HD 173667 & GJ9635 & F6V &2.79&19.1&4024320&No& \\
HD 181321 & GJ755 & G5V &4.89&20.9&4030208&No& \\
HD 185144 & & K0V &5.93&5.6&4024576&No& \\
HD 188376 & & G5V &2.82&23.8&4025088&No& \\
HD 191408 & GJ783A & K3V &6.41&6&4025600&No. New& \\
HD 196378 & GJ794.2 & F8V &3.2&24.2&4025856&No. New& \\
HD 203608 & $\gamma$ Pav & F8V &4.4&9.2&4026368&No& \\
HD 212330 & GJ857 & G3IV &3.76&20.5&4027136&No. New& \\
HD 216437$^5$  & $\rho$ Ind &   G2.5IV &6.06&26.5&4084224&No& \\
HD 217014$^5$  & 51 Peg&   G2.5IV&5.49&15.4&4027648&No. New& \\
HD 217813 & & G5 &4.72&24.3&4030464&No. New& \\
HD 222368 & & F7V &3.43&13.8&4028416&No. New& \\
HD 225239 & GJ3002 & G2V &3.28&36.8&4028672&N/A& \\
  {\it Follow-up}  &  {\it Observations} &&&&&& \\
HD 128311$^5$  & GJ 3860&   K0 &7.51&16.5&4083712&70 $\micron$& \\
HD 206860 & GJ836.7 & G0V &4.62&18.4&12719872&70 $\micron$& Long\\
\enddata
\tablenotetext{1}{Two other stars, HD 190248 (AOR 4025344) and HD
209100 (AOR 4026625), were observed as part of this program using
the IRS High Resolution mode. These stars will be discussed
elsewhere. HD 33262 was not observed properly due to a pointing error
and is not considered further.}
\tablenotetext{2}{Spectral Types from SIMBAD. Visual magnitudes and
distances as quoted in SIMBAD, typically from the Hipparcos
satellite (1997).}
\tablenotetext{3}{Star either does or does not have 24
or 70 $\mu$m excess as observed by MIPS. 'New' means MIPS data were obtained subsequently to data presented in Beichman et al 2005a or Bryden et al 2006 and are described in Table~\ref{newmipstable}. 'N/A' means star has not yet been observed with MIPS due to scheduling constraints.}
\tablenotetext{4}{Star has excess as observed by IRS as described in this paper at either Short (8-13 $\micron$) or Long (30-34 $\micron$) wavelengths.}
\tablenotetext{5}{Star has at least one radial velocity  planet.}
\end{deluxetable}

\clearpage

\begin{deluxetable}{l|rrrr|rrrr}
\tabletypesize{\scriptsize} 
\tablecaption{ IRS Excess in 8.5-13 and 30-34 $\mu$m Bands
\label{filtersummary}} 
\tablehead{ &Excess (mJy)&Frac. &Excess/&$\ld$($^1$)&Excess (mJy)&Frac. &Excess/&$\ld$($^1$)\\
Star&(8.5-13 $\micron$)&Excess&Dispersion$^2$&($10^{-5}$)&(30-34 $\micron$)&Excess&Dispersion$^2$&($10^{-5}$) } 
\startdata
HD 693& -20.8$\pm$1.3& -0.015& -1.2& $<$8.2& 0.3$\pm$0.8& 0.002& 0.1&$<$0.4 \\
HD 3795&-5.5$\pm$0.5& -0.008& -0.7& $<$13& -1.9$\pm$0.6& -0.026& -1.3&$<$0.7 \\
HD 4628&13.8$\pm$0.9& 0.009& 0.7& $<$26.8& -0.8$\pm$1.3& -0.004& -0.2&$<$1.4 \\
HD 7570& 4.4$\pm$0.5& 0.004& 0.3& $<$9.7& 10.7$\pm$0.9& 0.082& 4.3& 0.8 \\
HD 9826&-7.4$\pm$0.9&-0.003& -0.3& $<$9.7& -1.3$\pm$1.5& -0.004& -0.2&$<$0.5 \\
HD 10800& 8.8$\pm$0.3& 0.015& 1.2& $<$13& -1.6$\pm$0.8& -0.024& -1.1&$<$0.7 \\
HD 39091& 7.4$\pm$0.4& 0.010& 0.8& $<$13& 0.0$\pm$3.3& -0.006& -0.1&$<$0.7 \\
HD 43834& 0.1$\pm$0.6& -0.001& 0.0& $<$13.9& 0.5$\pm$0.9& 0.004& 0.2&$<$0.7 \\
HD 55575& 4.1$\pm$0.5& 0.006& 0.4& $<$11.1& 0.8$\pm$0.9& 0.008& 0.4&$<$0.6 \\
HD 58855& 3.1$\pm$0.5& 0.005& 0.4& $<$9.7& -0.4$\pm$0.9& -0.004& -0.2&$<$0.5 \\
HD 69830$^3$& 72.3$\pm$4.5& 0.099& 7.0&  41.6& 42$\pm$1& 0.50& 24.4&8.0 \\
HD 72905& 22.3$\pm$1.2& 0.028& 2.2& 9.5& 7.0$\pm$0.5& 0.083& 4.5& 1.1 \\
HD 75732& 14.4$\pm$0.5& 0.019& 1.5& $<$13.9& 1.5$\pm$0.9& 0.019& 0.9&$<$0.7 \\
HD 76151& 10.0$\pm$0.3& 0.018& 1.4& $<$13& 10.1$\pm$0.7& 0.151& 7.3& 2 \\
HD 84737&21.1$\pm$0.6&0.017& 1.4& $<$11.1& -0.3$\pm$1.1& -0.002& -0.1&$<$0.6 \\
HD 86728& -13.7$\pm$0.7& -0.012& -0.9& $<$13& 0.1$\pm$0.8& 0.002& 0.1&$<$0.7 \\
HD 95128&-28.1$\pm$0.9&-0.022& -1.7& $<$11.1& 0.6$\pm$1.3& 0.006& 0.3&$<$0.6 \\
HD 101501$^4$& -98.4$\pm$1.8&-0.071& -5.6& $<$13.9& 0.6$\pm$1.1& 0.005& 0.3&$<$0.7 \\
HD 120136&15.8$\pm$0.6&0.011& 0.8& $<$9.7& -1.2$\pm$1.6& -0.007& -0.3&$<$0.5 \\
HD 128311&-0.9$\pm$0.4& -0.003& -0.2& $<$16& 1.4$\pm$0.6& 0.037& 1.6&$<$0.8 \\
HD 133002& 1.8$\pm$0.6& 0.002& 0.1& $<$11.1& 2.9$\pm$1.0& 0.025& 1.3&$<$0.6 \\
HD 136064& -6.6$\pm$0.6&-0.007& -0.6& $<$9.7& 0.2$\pm$1.0& 0.002& 0.1&$<$0.5 \\
HD 142373&1.2$\pm$0.7&0.001& 0.0& $<$11.1& -3.5$\pm$2.0& -0.017& -0.8&$<$0.6 \\
HD 146233&-28.1$\pm$1.2&-0.032&-2.5&$<$13& -1.6$\pm$0.9& -0.018& -0.9&$<$0.7 \\
HD 154088&-5.3$\pm$0.3&-0.013&-1.0&$<$13.9&-0.6$\pm$1.1& -0.017& -0.6&$<$0.7 \\
HD 166620&-0.3$\pm$0.4&0.000&0.0& $<$26.8& -0.2$\pm$0.6& -0.002& -0.1&$<$1.4 \\
HD 168151& 15.7$\pm$0.8&0.016&1.3& $<$8.2& -0.9$\pm$0.8& -0.009& -0.5&$<$0.4 \\
HD 173667& 46.4$\pm$0.9& 0.023& 1.8& $<$9.7& 9.7$\pm$1.5& 0.043& 2.3&$<$0.5 \\
HD 181321& 0.3$\pm$0.4& 0.001& 0.1& $<$13& 0.0$\pm$0.6& 0.002& 0.1&$<$0.7 \\
HD 185144& -5.5$\pm$2.1& -0.003& -0.2& $<$16& 3.2$\pm$2.0& 0.011& 0.6&$<$0.8 \\
HD 188376& 12.8$\pm$1.1& 0.005&0.4& $<$13& -3.5$\pm$2.2& -0.013& -0.7&$<$0.7 \\
HD 191408& -8.7$\pm$1.0&-0.004&-0.3& $<$26.8& 4.4$\pm$1.8& 0.020& 1.0&$<$1.4 \\
HD 196378& -15.3$\pm$1.2&-0.012&-0.9& $<$9.7& 2.1$\pm$0.8& 0.016& 0.8&$<$0.5 \\
HD 203608&-38.2$\pm$0.8&-0.016&-1.3&$<$9.7&-5.2$\pm$1.4& -0.019& -1.0&$<$0.5 \\
HD 206860& -0.6$\pm$0.4& -0.001& -0.1& $<$11.1& 5.3$\pm$0.7& 0.085& 4.0& 0.9 \\
HD 212330& 16.7$\pm$0.8& 0.014&1.1& $<$13& -0.4$\pm$0.8& -0.004& -0.2&$<$0.7 \\
HD 216437& 1.0$\pm$0.4& 0.003& 0.2& $<$13& -1.0$\pm$0.3& -0.019& -1.0&$<$0.7 \\
HD 217014& 2.0$\pm$0.5& 0.002& 0.2& $<$13& -0.7$\pm$1.0& -0.008& -0.4&$<$0.7 \\
HD 217813& 0.5$\pm$0.3& 0.001& 0.1& $<$13& -1.8$\pm$1.2& -0.054& -1.4&$<$0.7 \\
HD 222368& -14.4$\pm$1.6&-0.007&-0.5& $<$9.7& 7.0$\pm$1.7& 0.025& 1.3&$<$0.5 \\
HD 225239&-8.6$\pm$0.5& -0.015& -1.2& $<$13& 0.4$\pm$0.6& 0.005& 0.3&$<$0.7\\
\ \ \ Average$^5$  &  & &  & 14$\pm$6 &  & &  & 0.74$\pm$0.31 \\
\enddata
\tablenotetext{1}{\, Limits on $\ld$ from using Eqn (2)  and three times the  dispersion in the fractional excess of the whole population.}
\tablenotetext{2}{ Excess divided by dispersion averaged over entire sample.}
\tablenotetext{3}{ Data on excess taken from Beichman et al 2005b.}
\tablenotetext{4}{Star shows significant flat field errors after re-calibration and is not considered further.}
\tablenotetext{5}{Average value of $\ld$ (3$\sigma$ limits or detections) after 2$\sigma$ rejection of outliers.}
\end{deluxetable}

\clearpage

\begin{deluxetable}{ll|ccc|cccc|c}
\tabletypesize{\scriptsize} 
\tablecaption{New MIPS Observations$^1$ \label{newmipstable}} 
\tablehead{ Star & AOR & $F_\nu$ (24 $\micron$)$^2$&Phot.$^3$ &Obs./Phot. & $F_\nu$ (70 $\micron$)& Phot. & Obs./Phot. & $\chi$70$^4$ & $L_d/L_*$ (70 $\micron$)\\ &	&(mJy)	&(mJy)	&	&(mJy)	&(mJy)		&	&		&}
\startdata
HD693	&4030976&254&265&0.96&37.7$\pm$5.6&30.0&1.26&1.37& $< 0.4\times10^{-5}$\\
HD9826$^5$&4033280&519&539&0.96&51.2$\pm$10.3&60.7&0.84&-0.92&$<0.4\times10^{-5}$\\
HD154088&4061696&87	&70	&1.24	&-0.1$\pm$6.6	&7.8	&	&-1.19	& $< 2.7\times10^{-5}$\\
HD191408&4055040&443	&435	&1.02	&46.7$\pm$12.2	&50.1	&0.93	&-0.28	& $< 1.5\times10^{-5}$\\
HD196378&4055808&227	&254	&0.90	&30.5$\pm$5.6	&28.9	&1.06	&0.29	& $< 0.4\times10^{-5}$\\
HD212330&4058368&235	&234	&1.01	&26.4$\pm$5.8	&26.5	&1.00	&-0.01	& $<0.7\times10^{-5}$\\
HD217014&4058880&188	&178	&1.06	&25.2$\pm$5.3	&20.0	&1.26	&1.00	& 	   $<0.2\times10^{-5}$\\
HD217813&4062464&59	&70	&0.85	&3.7$\pm$4.8	&7.9	&0.47	&-0.87	& $< 1.8\times10^{-5}$\\
HD222368&4060160&514	&534	&0.96	&54.$\pm$10.2	&60.4	&0.91	&-0.56	& $< 0.4\times10^{-5}$\\
Average& 	&	&	&1.00$\pm$0.04&	& &0.92$\pm$0.09& &\\
\enddata
\tablenotetext{1}{MIPS data obtained after data presented in Beichman et al 2005a and Bryden 2006}
\tablenotetext{2}{Statistical noise is negligible.}
\tablenotetext{3}{Photospheric prediction based on extrapolation from
2MASS and shorter wavelength data as described in the text and in Bryden et al. 2006.}
\tablenotetext{4}{$\chi$70 defined as $(Observed-Photosphere)/Noise$.}
\tablenotetext{5}{Ups And}
\end{deluxetable}

\clearpage

\begin{deluxetable}{lccc}
\tablecaption{MIPS Excesses \label{mipstable}}
\tablehead{ Star &$F_\nu$ (70 $\micron$)  Excess (mJy)$^1$ & $\ld$ (70 $\micron$) &  $\ld$ (10-70 $\micron$) } \startdata
HD 7570   &   14$\pm$6  &0.38$\times10^{-5}$  &0.43$\times10^{-5}$  \\
HD 69830$^2$  &   $<$3      &  $<0.5\times10^{-5} $ &20$\times10^{-5}$\\
HD 76151  &   16$\pm$4  &1.0$\times10^{-5}$  &1.2$\times10^{-5}$  \\
HD 72905$^3$ &   24$\pm$4  &1.2$\times10^{-5}$  &2.9$\times10^{-5}$  \\
HD 128311 &   18$\pm$4  &2.6$\times10^{-5}$  &  \\
HD 206860 &   15$\pm$4  &0.92$\times10^{-5}$  &15$\times10^{-5}$  \\\
\enddata
\tablenotetext{1}{Bryden et al 2006}
\tablenotetext{2}{Beichman et al 2005b}
\tablenotetext{3}{The star HD72905 has a $70\mu$m\ flux density of 24$\pm$3 mJy \citep{bryden2006}, a factor of 2-3 lower than measured by ISO-PHOT at 60 and 90 $\mu$m\ \citep{spangler2001}. We adopt the MIPS value.}
\end{deluxetable}

\clearpage

\begin{deluxetable}{l|cc|cccccc}
\tabletypesize{\scriptsize} 
\tablecaption{ Model Fits to Spectra \label{modtable}}
\tablehead{Star & R$_{BB}$ & T$_{BB}$ & R$_1$-R$_2$ & T$_1$-T$_2$ & Optical Depth  & Reduced & M$_{grain}$ & M$_{total}$      \\
&(AU)&(K)&(AU)&(K) & $\tau$ (10 $\mu$m)&$\chi^2$ &$\MEarth$ &$\MEarth$$<$10 km }
\startdata
HD 7570 & 7.8$\pm$0.8 & 98$\pm$20& 11-12 & 83-86 & 1.4$\times$10$^{-5}$ & 0.68$^1$  &1.3$\times$10$^{-6}$ & 0.042 \\
HD 76151& 8.1$\pm$0.6 & 90$\pm$14  & 7.6-8.7 & 93-86 & 21$\times$10$^{-5}$ & 0.62$^1$  & 1.6$\times$10$^{-6}$ & 0.052 \\
HD 128311&   $>15$ &$<55$  &  & &  &  &  &  \\
HD 206860& 13.3$\pm$1.2 & 73$\pm$13 & 13.2-14.3 & 74-71 & 21$\times$10$^{-5}$ &0.5$^1$  & 2.7$\times$10$^{-6}$ & 0.087 \\

\hline
HD 72905 & & &  &  &  & & & \\
\ \ \ 0.25 $\micron$ silicate & & & 0.03-0.43 & 1500-540 & 10$\times$10$^{-5}$ & & 7.2$\times$10$^{-10}$ & 1.4$\times$10$^{-4}$ \\
and & & & & & & \\
\hline
\ \ \ 0.25 $\micron$ silicate & & & 109-315 & 55-36&33 $\times10^{-6}$ &0.83$^2$  &
110$\times$10$^{-6}$ & 22 \\ 
or & & & & & & & & \\
\ \ \ 10 $\micron$  silicate & 14$\pm$1 & 70$\pm$10 & 12.2-15.9 & 67-63  & 7$\times$10$^{-5}$ & 0.83$^2$ & 3.3$\times$10$^{-6}$ & 0.10 \\
\enddata
\tablenotetext{1}{$\lambda>21 \mu$m, 74 degrees of freedom}
\tablenotetext{2}{$\lambda>8 \mu$m, 256 d.o.f.} 
\end{deluxetable}

\clearpage

\begin{figure}[ht]
\epsscale{1} \plotone{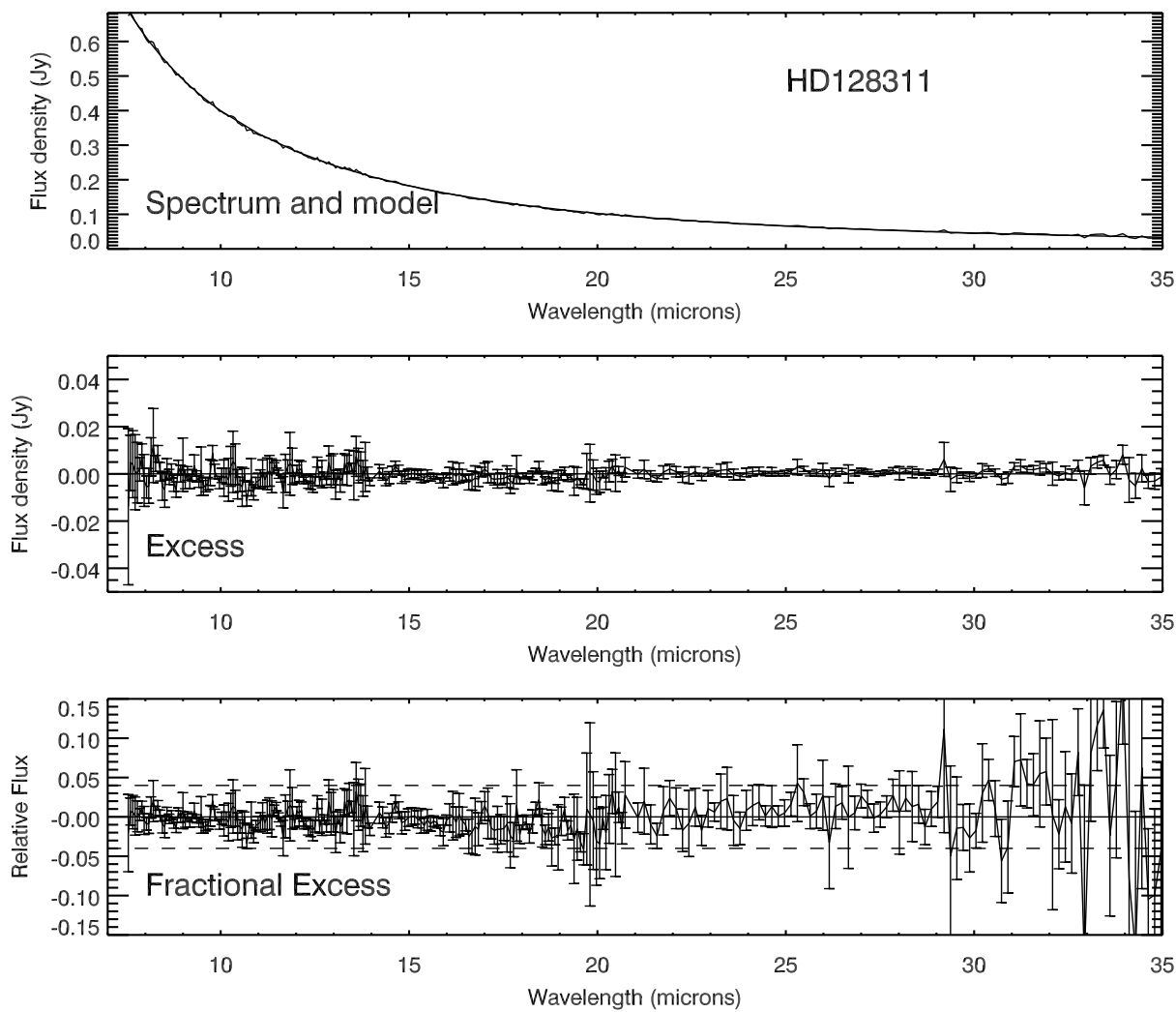}

\caption{These plots show the IRS spectrum (top), excess flux (middle), and fractional excess (bottom) as a function of wavelength for HD128311, HD 216437, HD 206860, HD7570, HD72905 and HD76151. The first two stars have {\it no measurable excess} and are shown as representative of the majority of the sample. Also plotted in the lower figure are the approximate 2$\sigma$ dispersion in the deviations from the photospheric 
models.  Fractional deviations between these lines should be regarded with skepticism.\label{three}}
\end{figure}
\begin{figure}[ht]
\epsscale{1} \plotone{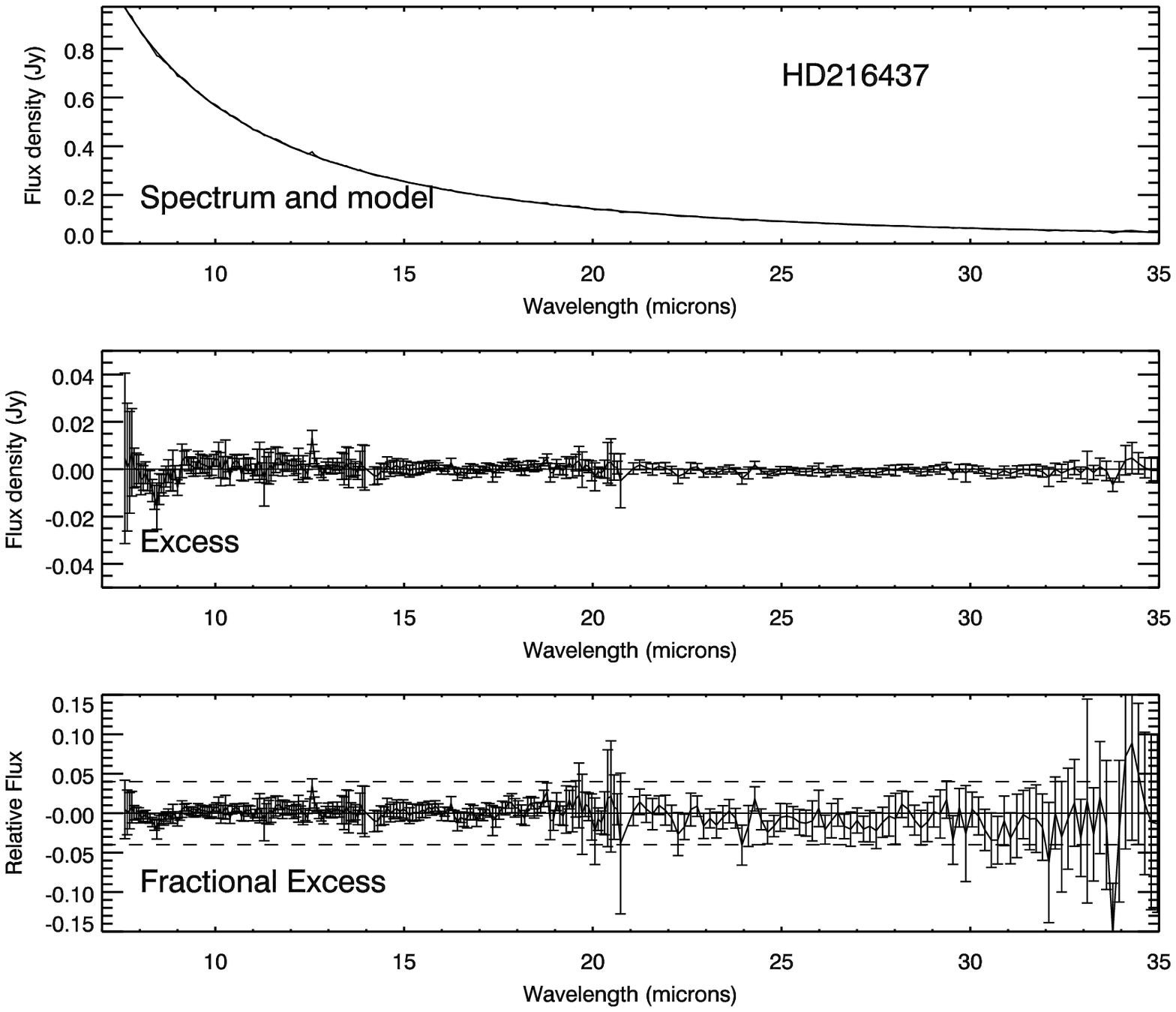}
\figurenum{\ref{three}}\caption{continued.}
\end{figure}

\begin{figure}[ht]
\epsscale{1} \plotone{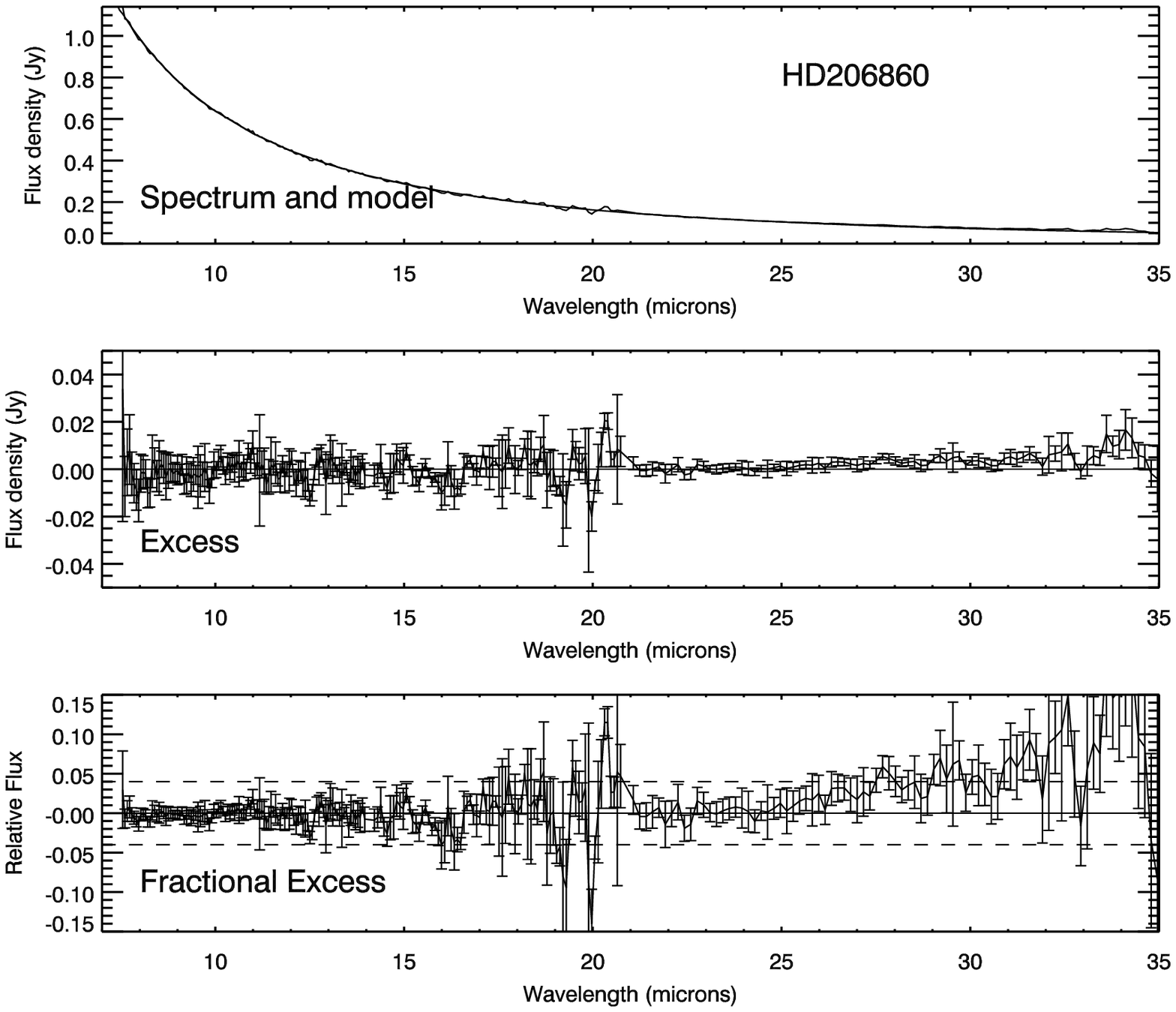}
\figurenum{\ref{three}}\caption{continued.}
\end{figure}
\begin{figure}[ht]
\epsscale{1} \plotone{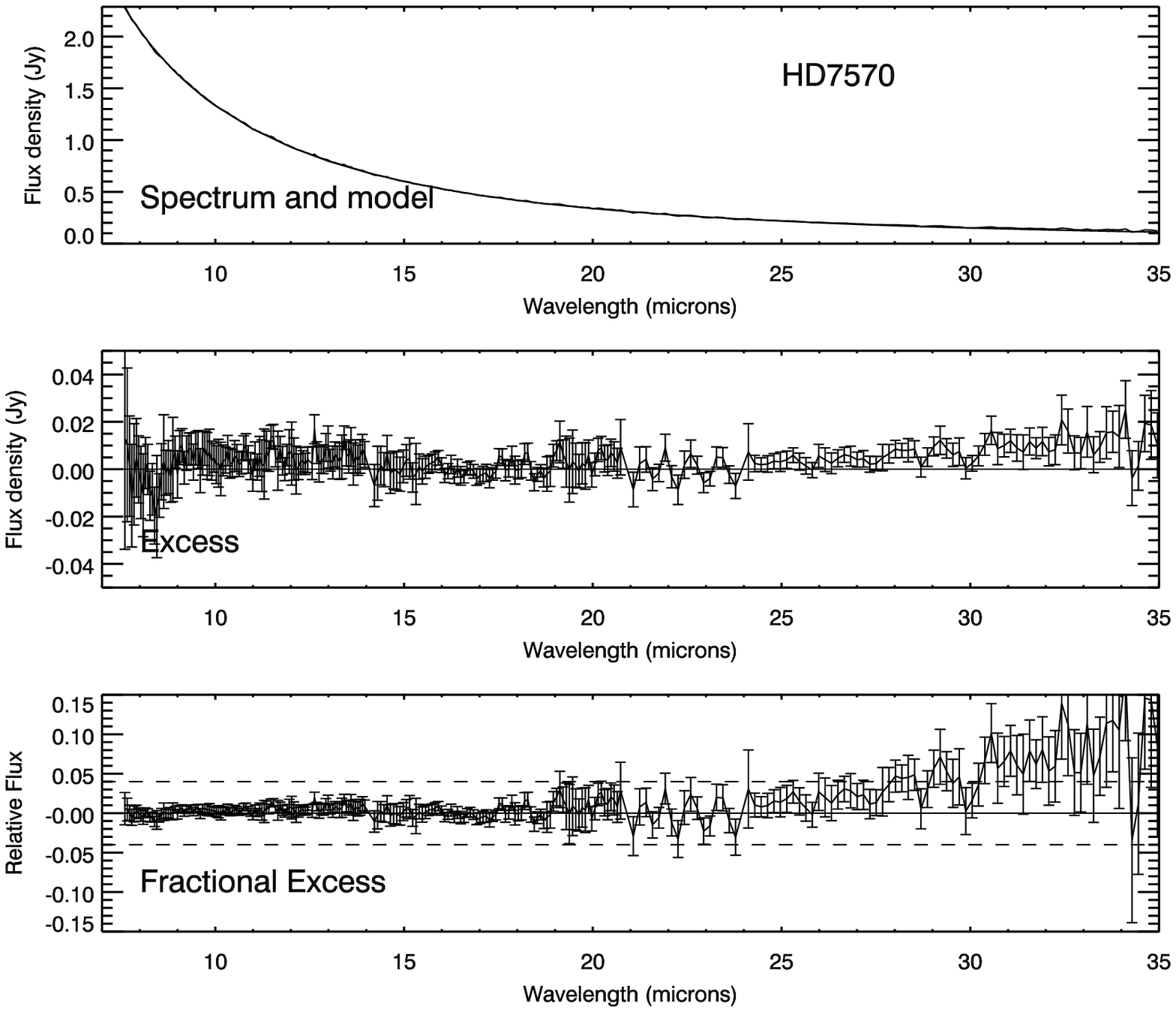}
\figurenum{\ref{three}}\caption{continued.}
\end{figure}

\begin{figure}[ht]
\epsscale{1} \plotone{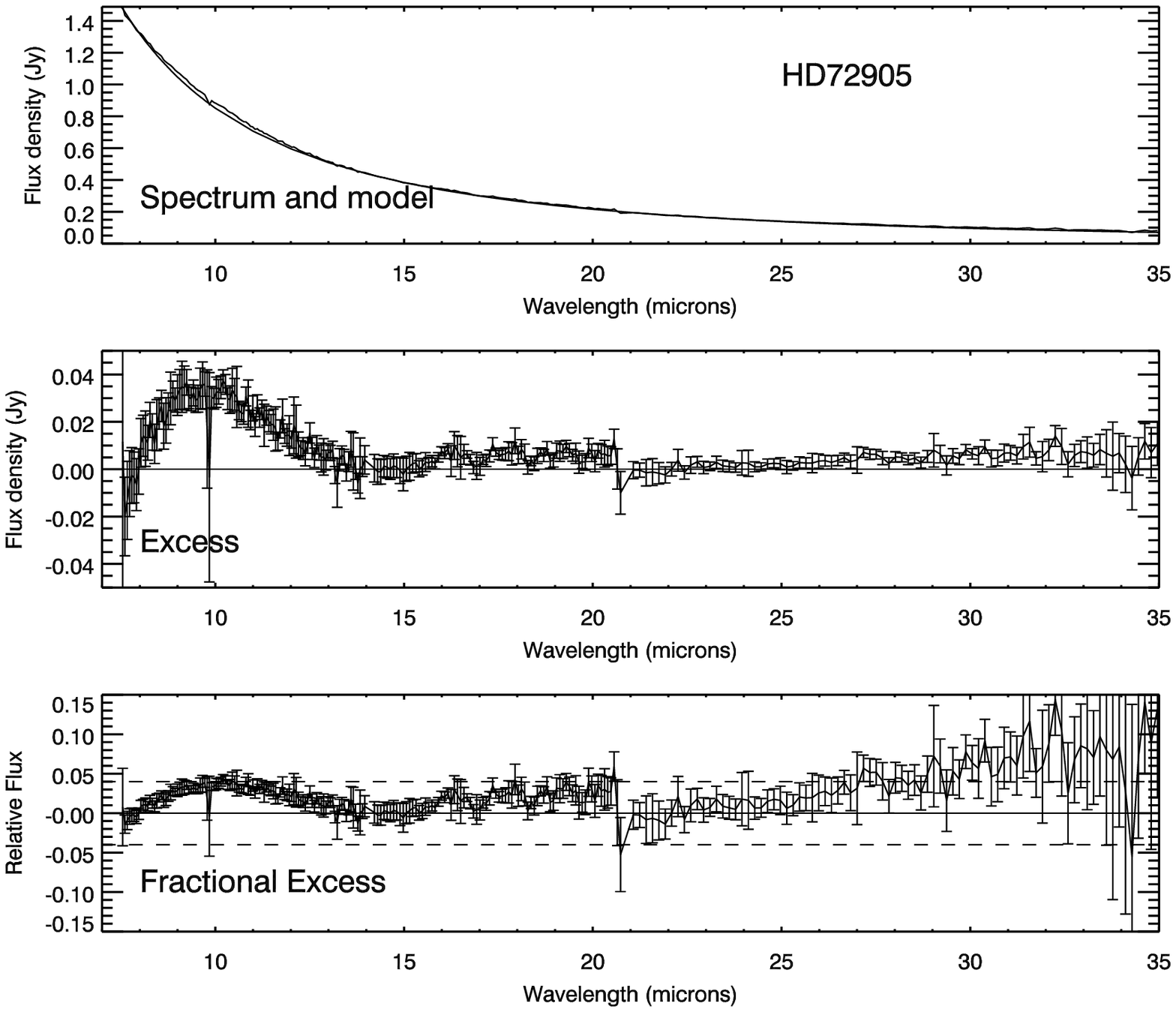}
\figurenum{\ref{three}}\caption{continued.}
\end{figure}
\begin{figure}[ht]
\epsscale{1} \plotone{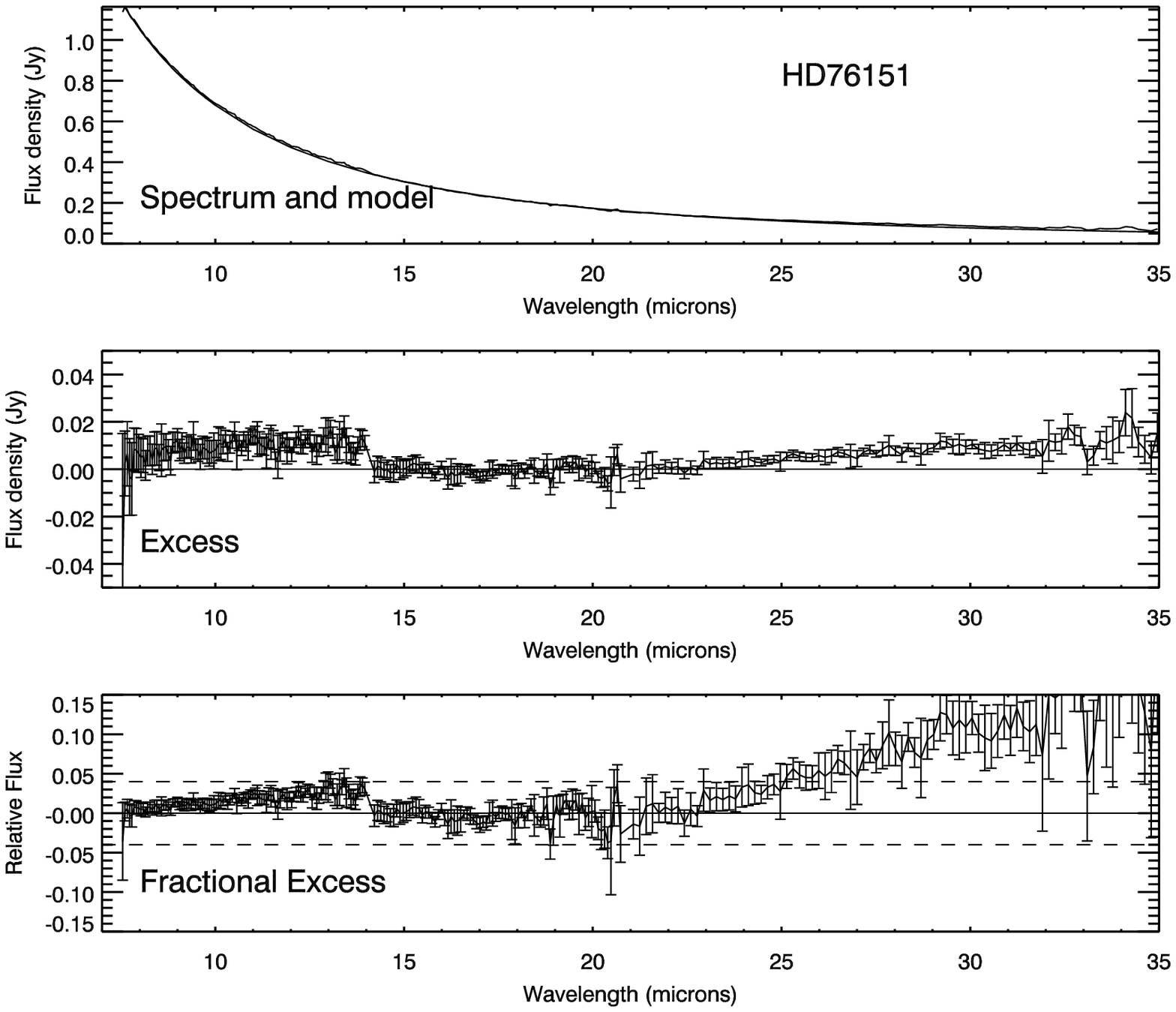} 
\figurenum{\ref{three}}\caption{continued.}
\end{figure}

\clearpage

\end{document}